\documentstyle[preprint,aps]{revtex}

\tighten
\begin{document}
%\draft

\title{
Large-basis shell-model calculation of $^{10}$C$
\rightarrow ^{10}$B Fermi matrix element
}
\medskip

\author{
        P. Navr\'atil$^a$\footnote{On leave of absence from the
   Institute of Nuclear Physics,
                   Academy of Sciences of the Czech Republic,
                   250 68 \v{R}e\v{z} near Prague,
                     Czech Republic.},
        B. R. Barrett$^a$, and W. E. Ormand$^b$
        }

\medskip

\address{
$^a$Department of Physics, University of Arizona, Tucson, 
Arizona 85721 \\
$^b$202 Nicholson Hall, Department of Physics and Astronomy,
Louisiana State University, Baton Rouge, LA 70803-4001
}

\maketitle

\bigskip

\begin{abstract}
We use a $4\hbar\Omega$ shell-model calculation
with a two-body effective interaction derived microscopically
from the Reid93 potential to calculate the isospin-mixing correction
for the $^{10}$C$\rightarrow ^{10}$B superallowed Fermi transition.
The effective interaction takes into account the Coulomb potential
as well as the charge dependence of $T=1$ partial waves.
Our results suggest the isospin-mixing correction 
$\delta_{\rm C}\approx 0.1\%$, which is compatible with previous
calculations. 
The correction obtained in those calculations, performed in
a $0\hbar\Omega$ space, was dominated by deviation from unity of
the radial overlap between the converted proton and the corresponding 
neutron. In the present calculation this effect is accommodated
by the large model space. 
The obtained $\delta_{\rm C}$ correction
is about a factor of four too small to obtain unitarity
of the Cabibbo-Kobayashi-Maskawa matrix with the present
experimental data.
\end{abstract}

\bigskip
\bigskip
\bigskip

\narrowtext

%\noindent{\small \bf PACS Numbers 21.60.Cs, 23.40.Hc, 27.20.+n}

%\noindent{\small {\bf Keyword abstract:}

\section{Introduction}
\label{sec1}

Superallowed Fermi $\beta$ transitions in nuclei,
$(J^\pi=0^+,T=1)\rightarrow (J^\pi=0^+,T=1)$,
provide an excellent laboratory for
precise tests of the properties of the electroweak interaction, 
and have been the subject of intense study for several decades (cf.
Refs.
~\cite{r:ref1,r:Tow77,r:OB89,r:Har90,r:Sir78,r:CKM,r:Bli73,r:Sir86,r:outer,r:Bar92,r:OB85,r:Tow89,r:OB95}. 
According to 
the conserved-vector-current (CVC) hypothesis, for pure Fermi transitions 
the product of the partial half-life, $t$, and the statistical phase-space 
factor, $f$, should be nucleus independent and given by
\begin{equation}
ft=\frac{K}{G_V^2|M_F|^2},
\end{equation}
where 
$K/(\hbar c)^6=2\pi^3\ln 2 \hbar /(m_ec^2)^5=
8.120270(12)\times 10^{-7}~{\rm GeV}^{-4}{\rm s}$, $G_V$
is the vector coupling constant for nuclear $\beta$
decay, and $M_F$ is the Fermi
matrix element, $M_F=\langle\psi_f\mid T_\pm \mid\psi_i\rangle$. By
comparing the decay rates for muon and nuclear Fermi $\beta$ decay, the 
Cabibbo-Kobayashi-Maskawa (CKM) mixing matrix element~\cite{r:CKM} between 
$u$ and $d$
quarks ($v_{ud}$) can be determined 
and a precise test of the unitarity
condition of the CKM matrix under the assumption of the three-generation
standard model is possible~\cite{r:Sir78,r:CKM}. 

For tests of the standard model, two
nucleus-dependent corrections must be applied to experimental $ft$
values. The first is a series of radiative corrections to the statistical
phase-space factor embodied in the factors $\delta_R$ and $\Delta_R$,
giving~\cite{r:Bli73,r:Sir86,r:outer}
\begin{equation}
f_R=f(1+\delta_R+\Delta_R),
\end{equation}
where $\delta_R$ 
is due to standard, electromagnetic (``inner'')
radiative corrections (cf. p.~45 in Ref.~\cite{r:Bli73}) and 
$\Delta_R$ is what has been referred to as the ``outer'' radiative
correction (cf. p. 47 of Ref.~\cite{r:Bli73}) and includes axial-vector 
interference terms~\cite{r:outer,r:Bar92}. The second correction,
which is the subject of this work, arises because of the presence
of isospin-nonconserving (INC) forces (predominantly Coulomb) in nuclei 
that lead to 
a renormalization of the Fermi matrix element. This correction is  
denoted by $\delta_C$~\cite{r:Tow77,r:OB89,r:Tow89} and modifies the Fermi
matrix element by $\mid M_F\mid^2=\mid M_{F0}\mid^2(1-\delta_C)$, where
$M_{F0}=[T(T+1)-T_{Z_i}T_{Z_f}]^{1/2}$ is the value of the matrix element
under the assumption of pure isospin symmetry. 

With the corrections $\delta_R$, $\Delta_R$, and $\delta_C$, 
a ``nucleus-independent'' ${\cal F}t$ can be defined by
\begin{equation}
{\cal F}t=ft(1+\delta_R+\Delta_R)(1-\delta_C),
\end{equation}
and the CKM matrix element $v_{ud}$ is given by~\cite{r:Bar92}
\begin{equation}
\mid v_{ud}\mid^2 = \frac{\pi^3\ln 2}{{\cal F}t}\frac{\hbar^7}{G_F^2m_e^5c^4}
=\frac{2984.38(6)~{\rm s}}{{\cal F}t},
\label{e:vud}
\end{equation}
where the Fermi coupling constant, $G_F$, is obtained from muon $\beta$-decay,
and includes radiative corrections.
Currently, $ft$ values for nine superallowed transitions have been measured
with an experimental precision of 0.2\% or better~\cite{r:Har90,r:Sav95}.
With these precise measurements and reliable estimates for the corrections,
the CVC hypothesis can be confirmed by checking the constancy of the 
${\cal F}t$ values for each nucleus, while the unitarity condition of the 
CKM matrix is tested by comparing the average value of $v_{ud}$ 
with the values determined for $v_{us}=0.2199(17)$~\cite{r:Bar92}  and 
$v_{ub}<0.0075$ (90\% confidence level)~\cite{r:Tho88}, i.e.,  
$v^2=v_{ud}^2+v_{us}^2+v_{ub}^2=1$.

In the past, the nuclear structure correction $\delta_C$ has been computed
within the framework of the nuclear 
shell model~\cite{r:Tow77,r:OB89,r:OB85,r:Tow89,r:OB95}. In general, the
isospin-nonconserving components of the nuclear Hamiltonian are small, 
and can be treated perturbatively. Due to computational limitations
and uncertainties associated with determining an effective Hamiltonian, 
almost all calculations for nuclei with $A\ge 10$ have been performed
within a single major oscillator shell, e.g., for $^{10}$C the model space 
spanned by the $0p_{3/2}$ and $0p_{1/2}$ oribitals ({\it p}-shell). 
Within this context, two types of isospin mixing must be accounted for. 
The first is due to the mixing between states that lie within the 
shell-model configuration space. For example, for $A=10$, there are 
2, 7, and 1 {\it p}-shell configurations leading to 
$J^\pi=0^+$ and $T=0$, 1, and 2, respectively. Because of its
two-body nature, the INC interaction is composed of isospin operators
of rank zero (isoscalar), one (isovector), and two (isotensor), and
in the case of $A=10$ it is capable of mixing together all $J^\pi=0^+$ 
states. Traditionally, the configuration mixing correction is denoted as 
$\delta_{IM}$ and in Ref.~\cite{r:OB85} it was shown that the best 
estimates for $\delta_{IM}$ are obtained using an INC interaction that 
correctly describes the Coulomb energy splittings of the binding energies 
between members of the isospin multiplet, e.g., the $J^\pi=0^+, T=1$ states
in $^{10}$C, $^{10}$B, and $^{10}$Be. Of the two types of mixing,
$\delta_{IM}$ is the smallest with a magnitude of approximately 0.04-0.1\%. 

In addition to the mixing between states contained within
the shell-model configuration space, mixing with states that lie outside
the model space must also be accounted for. In particular, the Coulomb
interaction can strongly mix one particle-one hole ($1p-1h$) $2\hbar\Omega$
excitations, e.g., $0p_{3/1}\rightarrow 1p_{3/2}$, into the ground state. 
In previous works, excitations of this type were accounted for by examining
differences in the single-particle radial wave functions. Indeed, for
closed-shell configurations, mixing with $1p-1h$ states is properly
accounted for at the level of Hartree-Fock. Hence, the second correction
to the Fermi matrix element, denoted by $\delta_{RO}$, was estimated by 
evaluating the mismatch in the radial overlap between the single-particle
wave functions of the converted proton and the corresponding neutron. 
The explicit details for the calculation of $\delta_{RO}$, which involve a 
sum over intermediate $A-1$ parent states that then determine the 
proton and neutron separation energy for the radial wave function, 
are given in Refs.~\cite{r:Tow77,r:OB85}. For the most part, $\delta_{RO}$
is found to be the larger of the two components 
(with $\delta_C=\delta_{RO}+\delta_{IM}$) and has a magnitude of the order
0.1-0.8\%.

At present, two methods for evaluating $\delta_{RO}$ are espoused. 
The first (THH)~\cite{r:Tow77} uses Woods-Saxon (WS) radial wave functions, 
while in the second (OB)~\cite{r:OB89,r:OB85,r:OB95}, 
Hartree-Fock (HF) wave functions are employed. Generally speaking, the two
methods yield approximately the same dependence on nucleon number $A$, but the
HF values are systematically smaller by 0.1\% for the magnitude of the 
correction. The reason for the difference lies
in the HF mean field. The principal effect of the Coulomb interaction is to
push the proton wave functions out relative to the neutrons, hence, providing
a mismatch in the radial overlap. In Hartree-Fock, however, the proton and
neutron mean fields are coupled, and the Coulomb interaction actually induces
an attractive isovector mean field between the protons and neutrons. In 
effect, the Coulomb interaction pushes the protons out, but because of 
the strong interaction, the protons pull the neutrons out with them, hence, 
reducing the magnitude of the radial overlap mismatch.

When all {\it known} corrections, i.e., $\delta_R$, $\Delta_R$, $\delta_{RO}$,
and $\delta_{IM}$, are applied to the nine experimental 
data~\cite{r:Har90,r:Sav95}, it is found that
the ${\cal F}t$ values are essentially constant within the limits of
uncertainty but the unitarity limit is violated at the level of approximately
0.4(1)\% or 0.3(1)\% for the OB and THH corrections, respectively. In 
addition, preliminary data from a new experiment for 
$^{10}$C\cite{r:Fuj96} leads to
an ${\cal F}t$ value that is significantly smaller than that of 
Ref.~\cite{r:Sav95}, and has been interpreted as possible evidence for an,
as yet, unaccounted for correction that might lead to satisfying the 
unitarity condition of the CKM matrix. In addition, it must be admitted that
the present separation between the configuration mixing and radial overlap
contributions to $\delta_C$ is somewhat unsatisfying. 
A much better approach would be to perform a shell-model calculation that
includes several $\hbar\Omega$ excitations, so that both corrections would
evaluated on the same footing and simultaneously. 
Because of recent improvements in computational
capabilities and the ability to determine an effective model-space Hamiltonian 
based on realistic nucleon-nucleon interactions, it is now possible to 
perform such a calculation for the lightest of the nine accurately measured
transitions. We report here the results of large-basis shell-model
calculations that include excitations up to $4\hbar\Omega$ for $A=10$ 
nuclides, with an emphasis on evaluating the isospin-mixing corrections
to the matrix element for the Fermi decay of $^{10}$C.

The organization for the paper is as follows. First, in 
Section~\ref{sec2} we discuss the shell-model Hamiltonian
with a bound center-of-mass, 
the method used to derive the starting-energy-independent
effective interaction, and the renormalization of the transfer
operator. 
Results of the Fermi matrix element calculations are presented in
section \ref{sec3} and 
concluding remarks are given in section \ref{sec4}.

\section{Shell-model Hamiltonian and the effective interaction}
\label{sec2}

In our calculation we use the one- plus two-body Hamiltonian
for the A-nucleon system, i.e.,
\begin{equation}\label{ham}
H=\sum_{i=1}^A \frac{\vec{p}_i^2}{2m}+\sum_{i<j}^A 
V_{\rm N}(\vec{r}_i-\vec{r}_j) \; ,
\end{equation}
where $m$ is the nucleon mass and $V_{\rm N}(\vec{r}_i-\vec{r}_j)$ 
the nucleon-nucleon interaction,
modified by adding the center-of-mass harmonic-oscillator potential
$\frac{1}{2}Am\Omega^2 \vec{R}^2$, 
$\vec{R}=\frac{1}{A}\sum_{i=1}^{A}\vec{r}_i$.
This potential does not influence intrinsic properties of the 
many-body system. It provides, however, a mean field felt by each nucleon
and allows us to work with a convenient harmonic-oscillator basis.
The modified Hamiltonian, depending on the harmonic-oscillator 
frequency $\Omega$, may be cast into the form
\begin{equation}\label{hamomega}
H^\Omega=\sum_{i=1}^A \left[ \frac{\vec{p}_i^2}{2m}
+\frac{1}{2}m\Omega^2 \vec{r}^2_i
\right] + \sum_{i<j}^A \left[ V_{\rm N}(\vec{r}_i-\vec{r}_j)
-\frac{m\Omega^2}{2A}
(\vec{r}_i-\vec{r}_j)^2
\right] \; .
\end{equation}
The one-body term of the Hamiltonian (\ref{hamomega}) is then re-written
as a sum of the center-of-mass term
$H^\Omega_{\rm cm}=\frac{\vec{P}_{\rm cm}^2}{2Am}
+\frac{1}{2}Am\Omega^2 \vec{R}^2$,
$\vec{P}_{\rm cm}=\sum_{i=1}^A \vec{p}_i$,
and a term depending on relative coordinates only.
Shell-model calculations are carried out in a model space defined
by a projector $P$. In the present work, we will always use a complete 
$N\hbar\Omega$ model space. The complementary space to the model space
is defined by the projector $Q=1-P$.
In addition, from among the eigenstates of the Hamiltonian 
(\ref{hamomega}),
it is necessary to choose only those corresponding to the same 
center-of-mass energy. This can be achieved by projecting 
the center-of-mass eigenstates
with energies greater than $\frac{3}{2}\hbar\Omega$ upwards in the
energy spectrum. The shell-model Hamiltonian, used in the actual 
calculations, takes the form
\begin{eqnarray}\label{phamomegabeta}
H^\Omega_{P\beta}=\sum_{i<j=1}^A &P&\left[ 
\frac{(\vec{p}_i-\vec{p}_j)^2}{2Am}
+\frac{m\Omega^2}{2A} (\vec{r}_i-\vec{r}_j)^2
\right]P + \sum_{i<j}^A P\left[ V_{ij}-\frac{m\Omega^2}{2A}
(\vec{r}_i-\vec{r}_j)^2
\right]_{\rm eff} P  \nonumber \\
&+& \beta P(H^\Omega_{\rm cm}-\frac{3}{2}\hbar\Omega)P\; ,
\end{eqnarray}
where $\beta$ is a sufficiently large positive parameter.

The effective interaction introduced in Eq. (\ref{phamomegabeta})
should, in principle, exactly reproduce the full-space results
in the model space for some subset of states.
In practice, the effective interactions can
never be calculated exactly as, in general, for an A-nucleon system 
an A-body effective interaction is required. 
Consequently, large model spaces
are desirable when only an approximate effective interaction is used. 
In that case, the calculation
should be less affected by any imprecision of the effective
interaction. 
The same is true for the evaluation of any observable characterized
by an operator. In the model space, renormalized effective operators 
are required. The larger the model space, the less renormalization
is needed.

Usually, the effective Hamiltonian is approximated by a 
two-body effective interaction determined from a two-nucleon
system. In this study, we use the procedure as described in 
Ref.~\cite{r:NB96}.   
To construct the effective interaction we employ
the Lee-Suzuki \cite{r:LS80} similarity transformation
method, which gives an interaction in the form
$P_2 V_{\rm eff}P_2 = P_2 V P_2 + P_2V Q_2\omega P_2$,
with $\omega$ the transformation operator satisfying $\omega=Q_2 \omega P_2$.
The projection operators $P_2, Q_2=1-P_2$ project on the two-nucleon
model and complementary space, respectively.
Our calculations start with exact solutions of the Hamiltonian
\begin{equation}\label{hamomega2}
H^\Omega_2\equiv H^\Omega_{02}+V_2^\Omega=
\frac{\vec{p}_1^2+\vec{p}_2^2}{2m}
+\frac{1}{2}m\Omega^2 (\vec{r}^2_1+\vec{r}^2_2)
+ V(\vec{r}_1-\vec{r}_2)-\frac{m\Omega^2}{2A}(\vec{r}_1-\vec{r}_2)^2 \; .
\end{equation}
which is the shell-model Hamiltonian (\ref{hamomega}) applied to
a two-nucleon system. 
We construct the effective interaction directly
from these solutions. Let us denote the relative-coordinate two-nucleon 
harmonic-oscillator states, which form the model space, 
as $|\alpha_P\rangle$,
and those which belong to the Q-space, as $|\alpha_Q\rangle$.
Then the Q-space components of the eigenvector $|k\rangle$ of
the Hamiltonian (\ref{hamomega2}) can be expressed as a combination
of the P-space components with the help of the operator $\omega$
\begin{equation}\label{eigomega}  
\langle\alpha_Q|k\rangle=\sum_{\alpha_P}
\langle\alpha_Q|\omega|\alpha_P\rangle \langle\alpha_P|k\rangle \; .
\end{equation}
If the dimension of the model space is $d_P$, we may choose a set
${\cal K}$ of $d_P$ eigenevectors, 
for which the relation (\ref{eigomega}) 
will be satisfied. Under the condition that the $d_P\times d_P$ 
matrix $\langle\alpha_P|k\rangle$ for $|k\rangle\in{\cal K}$
is invertible, the operator $\omega$ can be determined from 
(\ref{eigomega}).  In the present application we select the lowest states
obtained in each channel. 
Once the operator $\omega$ is determined the effective Hamiltonian
can be constructed as follows 
\begin{equation}\label{effomega}
\langle \gamma_P|H_{2\rm eff}|\alpha_P\rangle =\sum_{k\in{\cal K}}
\left[
\langle\gamma_P|k\rangle E_k\langle k|\alpha_P\rangle
+\sum_{\alpha_Q}\langle\gamma_P|k\rangle E_k\langle k|\alpha_Q\rangle
\langle\alpha_Q |\omega|\alpha_P\rangle\right] \; .
\end{equation}
This Hamiltonian, when diagonalized in a model-space basis, reproduces
exactly the set ${\cal K}$ of $d_P$ eigenvalues $E_k$. Note that
the effective Hamiltonian is, in general, quasi-Hermitian. 
It can be hermitized by a similarity transformation 
determined from the metric operator $P_2(1+\omega^\dagger\omega)P_2$. 
The Hermitian Hamiltonian is then given by \cite{r:S82SO83}
\begin{equation}\label{hermeffomega}
\bar{H}_{\rm 2eff}
=\left[P_2(1+\omega^\dagger\omega)P_2\right]^{1/2}
H_{\rm 2eff}\left[P_2(1+\omega^\dagger\omega)
P_2\right]^{-1/2} \; .
\end{equation}

Finally, the two-body effective interaction used 
in the present calculations
is determined from the two-nucleon effective Hamiltonian 
(\ref{hermeffomega}) as $V_{\rm eff}=\bar{H}_{\rm 2eff}-H_{02}$.
Note that we distinguish the two-nucleon system projection operators
$P_2, Q_2$ from the A-nucleon system operators $P, Q$.

To at least partially take into account the many-body effects
neglected when using only a two-body effective interaction,
we employ the recently introduced multi-valued effective
interaction approach \cite{r:ZBVHS}. As a consequence, 
different effective interactions are used 
for different $\hbar\Omega$ excitations.
The effective interactions then carry an additional index 
indicating the sum of the oscillator quanta for the spectators,
$N_{\rm sps}$, defined by
\begin{equation}\label{Nsps}
N_{\rm sps} = N_{\rm sum} - N_{\alpha} - N_{\rm spsmin}
= N'_{\rm sum} - N_{\gamma} - N_{\rm spsmin} \; ,
\end{equation}
where $N_{\rm sum}$ and $N'_{\rm sum}$ are the total oscillator
quanta in the initial and final many-body states, respectively, 
and $N_{\alpha}$
and $N_{\gamma}$ are the total oscillator quanta in the initial
and final two-nucleon states $|\alpha\rangle$ and $|\gamma\rangle$,
respectively. $N_{\rm spsmin}$ is the minimal value of the spectator
harmonic-oscillator quanta for a given system. Here, for A=10, 
$N_{\rm spsmin}=4$.
Different sets of the effective interaction are determined
for different model spaces characterized by $N_{\rm sps}$ 
and defined by projection operators
\begin{mathletters}\label{projop}\begin{eqnarray}
Q_2(N_{\rm sps})&=&\left\{ 
\begin{array}{ll}
0 &  \mbox{if  $N_1+N_2\leq N_{\rm max} - N_{\rm sps}$} \; , \\ 
1 &  \mbox{otherwise} \; ;
\end{array}
\right.   \\
P_2(N_{\rm sps}) &=&
1-Q_2(N_{\rm sps}) \; .
\end{eqnarray}\end{mathletters}
In Eqs. (\ref{projop}), $N_{\rm max}$ characterizes the
two-nucleon model space. It is an input parameter chosen in relation
to the size of the many-nucleon model space.
This multi-valued effective-interaction approach is superior
to the traditional single-valued effective interaction,
as confirmed also in a model calculation \cite{r:NB96l}. 

Our goal in this study is to evaluate the Fermi matrix element
\begin{equation}\label{fermime}
M_{\rm F} = \langle ^{10}{\rm B}, 0^+1|T_-|^{10}{\rm C}, 0^+1\rangle \; ,
\end{equation}  
which is equal to $\sqrt{2}$ for an isospin-invariant system.
Note that for a system with isospin breaking, the isospin-lowering
operator $T_-$ should be renormalized in a similar way, as
the interaction used for calculation of the eigenstates appearing
in Eq. (\ref{fermime}). In fact, we can apply the formalism described
in Refs. \cite{r:effoper} to construct a two-body effective operator 
$(T_-)_{\rm eff}$
consistent with the two-body effective interaction derived above and exact
for the two-nucleon system. Then we could use such an operator in the A-body 
calculation. We studied such a possibility in a solvable-model calculation 
as described in Refs. \cite{r:NB96l}. Here, we did two-nucleon calculations 
with the effective $T_-$ operator. The observed renormalization of the
bare operator for the model spaces
of the size used in our calculations, was, however, insignificant
compared to the other effects as described further. 
Therefore, in the A-body calculations we used the bare $T_-$ operator.

\section{Application to the A=10 system with isospin breaking}
\label{sec3}

In order to evaluate the Fermi matrix element [Eq.~(\ref{fermime})],  we apply the 
formalism
outlined in section \ref{sec2} for A=10 nuclei. 
%$^{10}$C$\rightarrow ^{10}$B Fermi matrix element.
In the calculations we use the Reid93 nucleon-nucleon potential \cite{r:SKTS}
and consider the following isospin-breaking contributions. First, the Reid93
potential differs in the $T=1$ channels for pn and pp, nn systems, respectively.
Second, we add the Coulomb potential to the pp Reid93 potential. Consequently,
using the Eqs. (\ref{eigomega})-(\ref{hermeffomega}),
%,\ref{effomega},
we derive different two-body effective interactions for the pn, pp, and
nn systems. No other mechanisms for isospin breaking are considered.

As we derive the effective interaction microscopically from the nucleon-nucleon
interaction, the number of freely adjustable  parameters in the calculation 
is limited.

First, we have the choice of the model-space size in the shell-model 
diagonalization.
That is, however, constrained by computer capabilities. The largest model space
we were able to use was the space allowing all $4\hbar\Omega$ excitations 
relative
to the unperturbed ground state. Most of the calculations were done in the 
m-scheme using the Many-Fermion-Dynamics
Code \cite{r:VZ94} extended to allow the use of different pn, pp, nn 
interactions.
We also performed some calculations with the OXBASH 
shell-model code~\cite{r:oxbash}.
In the m-scheme, the dimensions associated with $^{10}$B and $^{10}$C are
581,740 and 430,137, respectively.
To study the dependence on the model-space size, we performed calculations
in the $2\hbar\Omega$ space as well. In that space, the dimensions drop 
to 14,502 and 10,111, respectively.   

Second, we have the choice of the two-nucleon model space used for
the evaluation of the effective interaction. This is related to the 
many-nucleon model-space size, and, in principle, is determined by that size.
Traditionally, however, the $Q=0$ space used to determine the G-matrix
does not, necessarily, coincide with the many-particle model space 
\cite{r:BHM71,r:HJKO95}.
In our calculation, the two-nucleon model space is characterized by 
a restriction on the number harmonic-oscillator quanta   
$N_1\le N_{\rm max}$, $N_2\le N_{\rm max}$, $(N_1+N_2)\le N_{\rm max}$.
Here, $N_i=2n_i+l_i$ is the harmonic-oscillator quantum number for the nucleon
$i, i=1,2$. This type of restriction guarantees an orthogonal transformation
between the two-particle states and the relative- and center-of-mass-coordinate
states. 
For the present $4\hbar\Omega$ calculation, 
the choice of $N_{\rm max}=6$ appears to be appropriate.
However, it has been observed in the past~\cite{r:NB96,r:ZBVC,r:ZBVM} that 
when the Lee-Suzuki
procedure combined with the G-matrix calculation according to 
Ref.~\cite{r:BHM71} 
(which is equivalent to the procedure we are using)
is applied to calculate the two-body effective interaction,
the resulting interaction may be too strong. This is, in particular, true,
when the multi-valued approach is used.    
Several possible adjustments were discussed to deal with this problem 
\cite{r:NB96,r:ZBVC}
and amounted to introducing an extra parameter. In the present calculations,
we do not introduce any new parameter, but rather
we treat $N_{\rm max}$ as a free parameter and use $N_{\rm max}=8$
for the $4\hbar\Omega$ calculations and $N_{\rm max}=6$ 
for the $2\hbar\Omega$ calculations, respectively. With this choice we obtain
quite reasonable binding energies for the studied nuclei.
We have also performed several $2\hbar\Omega$ 
test calculations with
single-valued interactions that were derived following  
Ref.\cite{r:NBthree}, 
as opposed to the multi-valued interaction 
discussed in the previous section. 
To obtain reasonable binding energies with the single-valued 
interaction we do not have to change the $N_{\rm max}$ value from that
corresponding to the many-nucleon space, e.g., $N_{\rm max}=4$ 
for the $2\hbar\Omega$ calculation. This difference of treatment of
the two types of interactions follows from the fact that the overall
strength of the single-valued interaction is weaker.

Third, our results depend on the harmonic-oscillator frequency $\Omega$.
We have studied this dependence by performing calculations
for the values $\hbar\Omega=14, 15.5$, and $17$ MeV.

Let us also mention one important feature of the present approach.
For both the multi-valued and the single-valued interactions our calculations
do not break the separation of the center-of-mass and the internal 
relative motion.
In particular, a variation of the parameter $\beta$ introduced in 
Eq. (\ref{phamomegabeta})
does not change the eigenenergies and other characteristic of the 
physical states.
This is so due to the choice of a complete $N\hbar\Omega$ many-nucleon space and
the triangular two-nucleon model space for deriving the effective 
interaction as well as the due to the procedure used to derive the effective
interaction.

In Figs. \ref{fig:B1014}, \ref{fig:B10155}, and \ref{fig:B1017} we present the
experimental and calculated spectra of $^{10}$B for 
$\hbar\Omega=14, 15.5$, and $17$ MeV, respectively, for the $2\hbar\Omega$
and $4\hbar\Omega$ model spaces. In general, we observe an overall 
improvement in the spectra
with the enlargement of the model space in all three cases. Also,
the $4\hbar\Omega$ calculations exhibit more stability with regard to 
changes in the harmonic-oscillator
frequency than do the $2\hbar\Omega$ results. The agreement with experiment
improves when going from $\hbar\Omega=14$ MeV to $\hbar\Omega=17$ MeV;
in particular for the ground state and the lowest states. In fact, 
from Fig.~\ref{fig:B1017} we find that a very reasonable
description of the spectra is obtained for $\hbar\Omega=17$ MeV. 

In Table \ref{tab:tab1}, the overall behavior with respect to $\hbar\Omega$ is
illustrated. In general, we observe a reasonable reproduction of the 
binding energy, with a moderate decrease occuring for increasing
$\hbar\Omega$. Using free-nucleon effective charges, 
we find that although the quadrupole moment for the $3^+0$ state is 
underestimated considerably, 
the magnetic dipole moment is well reproduced. 
In addtion, the point-proton rms
radius exhibits a fairly strong dependence, and increases with
decreasing $\hbar\Omega$. For the rms radius, we find that 
the best agreement with
experiment~\cite{r:OT96} is achieved for $\hbar\Omega=14$ MeV.

From the point of view of the beta decay of $^{10}$C, 
a good description of the $T=1$ states is important.
From Figs. \ref{fig:B1014}-\ref{fig:B1017} we can see that the calculated
$^{10}$B $T=1$ states have the right relative positions and are
reasonably stable with variations of both the 
model-space size and $\hbar\Omega$.
We have also performed $4\hbar\Omega$ calculations for $^{10}$Be to study
the splitting of the isospin analog states in the whole isospin-multiplet
$^{10}$C - $^{10}$B - $^{10}$Be. The experimental ground state splitting
between $^{10}$C and $^{10}$Be is 4.66 MeV, while our calculated values
are 4.68, 4.83, and 4.94 MeV for $\hbar\Omega=14, 15.5$, and $17$ MeV, 
respectively. The best agreement with experiment is achieved  
for $\hbar\Omega=14$~MeV, where the calculated
rms point-proton radius is also in agreement with the experimental value.
On the other hand,
the splitting between the $0^+1$ states of $^{10}$C and $^{10}$B,
which is experimentally 2.69 MeV, is overestimated in our calculations
by $8, 11$, and $14\%$ for the $\hbar\Omega=14, 15.$5, and $17$ MeV 
calculations, respectively. 
Since the correct $^{10}$C - $^{10}$Be splitting is obtained for
$\hbar\Omega=14$ MeV, 
the excess in the $^{10}$C - $^{10}$B splitting suggests that the
isospin breaking due the strong $T=1$ force may be too large.    
One possible explanation is that our approach for deriving the 
effective interaction tends to exaggerate
the differences between the pn and nn, pp potentials. Such an artificial
effect should decrease with increasing model-space size.
On the other hand, it is also possible that the Reid93 
potential itself overestimates differences between the pn and 
pp, nn systems in the $T=1$ channel. For the most part, we find the
best overall agreement for the rms point proton radius, binding energy,
and Coulomb energy splitting for $\hbar\Omega=14$ MeV. Given that the isospin
mixing is largely driven by the Coulomb interaction, which is then 
dependent on the size of the nucleus, we feel that the best
value for the isospin-mixing correction to the Fermi matrix element
will be achieved for $\hbar\Omega=14$ MeV.

The most important results of our study are also summarized in 
Table~\ref{tab:tab1} in the last two lines.
The calculated isospin-mixing corrections 
$\delta_{\rm C}=1-\frac{|M_{\rm F}|^2}{2}$, in \%, are presented for all
three choices of $\Omega$ and for both $4\hbar\Omega$ and $2\hbar\Omega$
model spaces. Again, a correlation between the radius and the isospin-mixing
correction is clearly observed, as $\delta_C$ decreases with increasing
radius. This is simply understood in terms of a larger radius 
implying weaker Coulomb effects. On the other hand, 
with an increase in the model-space size, a significant increase 
in the isospin-mixing correction is apparent. This is due to the
fact that in the larger model space, the excitation energies of the
1p-1h $0^+1$ states decrease, hence leading to greater mixing.
For this reason, the more realistic multi-valued effective interaction 
is important. We have also performed test calculations with the 
single-valued interaction in the $2\hbar\Omega$ space, and found 
$\delta_{\rm C}$ to be smaller by approximately 30\%.

Our $4\hbar\Omega$ results suggest an isospin-mixing correction
$\delta_{\rm C}\approx 0.08-0.1\%$. This is compatible with the previously
published value of $\delta_{\rm C}\approx 0.15(9)\%$ 
by Ormand and Brown \cite{r:OB95}. That value was a sum
of two contributions. First, about 0.04\% came from the shell-model
wave-function renormalization due to the isospin mixing and was obtained
in a $0\hbar\Omega$ shell-model calculation using phenomenological
effective interactions. Second, the amount 0.09\% was due to the deviation
from unity of the radial overlap between the converted proton and the
corresponding neutron. This effect was attributed to the influence
of states lying outside the $0\hbar\Omega$ space. The radial wave-functions
were obtained in a Hartree-Fock calculation using Skyrme-type interactions.
Because we use a multi-configuration model space in the present calculation,
we should have both effects included consistently at the same time.

Another important factor in the calculation is the position of the
$2\hbar\Omega$ states. As discussed before, the position of the
1p-1h states influences the ground-state isospin mixing. 
Unfortunately, the excitation energy of these states is not
known experimentally. However, in our calculations the multi-valued
effective interaction is used and a more realistic
description of these states should be obtained; 
especially, in the $4\hbar\Omega$ model space. 
On the other hand, in an analogous calculation
for $^4$He, it was observed that an $8\hbar\Omega$ model space
is needed to get the $2\hbar\Omega$ dominated $0^+$ state close
to the experimental excitation energy \cite{r:NB96,r:ZBVHS}.
There are states like $1^+0$ at 5.18 MeV in $^{10}$B or
$0^+1$ at 6.18 MeV in $^{10}$Be, which are believed to be 
2p-2h, $2\hbar\Omega$ excitations. We do not observe any 
such states below 7.5 and 12 MeV, respectively, in our calculations. 
The first excited $0^+1$ state in $^{10}$Be obtained in the 
$4\hbar\Omega$ calculation
with $\hbar\Omega=14$ MeV lies at 9.8 MeV. It is, however, 
predominately a $0\hbar\Omega$ state. 
There can be two reasons why we do not get such
states. First, these states have not yet converged in the Lanczos 
procedure. Second, and more likely, 
the $4\hbar\Omega$ model space is too small for the right description 
of the $2\hbar\Omega$ excitation states. 
Therefore, it would be desirable to extend the present calculations
to a larger, e.g., $6\hbar\Omega$ model space. Unfortunately, due to the 
computational limitations, it is not possible at this time to perform
a calculation of this magnitude. However, from Table \ref{tab:tab1}
we observe an increase of $\delta_{\rm C}$ by $\approx 0.03\%$ between
the $2\hbar\Omega$ and the $4\hbar\Omega$ calculation. Therefore, we 
might expect an increase of similar magnitude for an 
increase of the model-space
size beyond $4\hbar\Omega$. Therefore, the more realistic 
value of the isospin-mixing correction from our calculation 
would be $\delta_{\rm C}\approx 0.12(3)\%$, where the uncertainty is
estimated from the change in $\delta_C$ obtained when using an increased 
model space.

\section{Conclusions}
\label{sec4}

The effects of isospin mixing on the transition matrix element for the
superallowed Fermi $\beta$-decay of $^{10}$C were estimated within the 
context of a large-basis, shell-model calculation. The calculations
were performed assuming no closed core and an effective interaction
based on a realistic two-body nucleon-nucleon interaction, while including
the Coulomb interaction between protons. 
Contrary to previous estimates for the isospin corrections, this 
calculation was carried out
within a model space that included many $\hbar\Omega$ excitations.
As a consequence, the conventional configuration mixing and radial mismatch
contributions were evaluated within a unified framework, simultaneously and
the usual separation was not necessary. With regard to parameters used
within the calculation, we find a correlation between the isospin-mixing
correction and the Coulomb splitting between the isotopic multiplets, which,
in turn, is governed by the nuclear size through the oscillator parameter.
Given that the isospin-mixing correction is primarily a Coulomb effect,
the best value for $\delta_C$ is taken to coincide with the oscillator
parameter that correctly reproduces the Coulomb splittings. With regard
to the model-space size, a clear improvement (or an indication 
towards convergence) in most 
observables is evident when the size of the model space is increased 
from $2\hbar\Omega$ to $4\hbar\Omega$, but $\delta_C$ is found to increase
by only 0.03\% (in magnitude) in this case. Hence, our final estimate
for $\delta_C$ is taken to be 0.12(3)\% (where the $4\hbar\Omega$ result 
has been increased
by 0.03\% to account for the possible effects of an increased model space).
This result also happens to be in excellent agreement with the previous
estimates that relied on the conventional separation of the configuration and
radial mismatch contributions. Finally, we note that the magnitude of the
isospin-mixing correction obtained in our calculation does not lead to 
a resolution to the deviation from unitarity for the Cabibbo-Kobayashi-Maskawa
matrix.

\acknowledgements{
This work was supported by the NSF grant No. PHY96-05192.
P.N. also acknowledges partial support from 
the Czech Republic grant GA ASCR A1048504. W.E.O. acknowledges support 
from 
NSF Cooperative agreement No. EPS~9550481, NSF Grant No. 9603006, and DOE 
contract DE--FG02--96ER40985. 
Any opinions, findings, and conclusions or recommendations expressed
in this material are those of the authors and do not necessarily
reflect the views of the National Science Foundation.
}

\begin{figure}
\caption{The experimental and calculated excitation spectra of $^{10}$B.
The results corresponding to the model-space sizes
of $4\hbar\Omega$ and $2\hbar\Omega$ relative to the 
ground-state configurations are presented, respectively.
The harmonic-oscillator energy of $\hbar\Omega=14$ MeV was used.
}
\label{fig:B1014}
\end{figure}

\begin{figure}
\caption{
The same as in Fig.~\protect{\ref{fig:B1014}} for the
harmonic-oscillator energy of $\hbar\Omega=15.5$ MeV.
}
\label{fig:B10155}
\end{figure}

\begin{figure}
\caption{
The same as in Fig.~\protect{\ref{fig:B1014}} for the
harmonic-oscillator energy of $\hbar\Omega=17$ MeV.
}
\label{fig:B1017}
\end{figure}

\begin{table}
\begin{tabular}{ccccc}
&Exp&$\hbar\Omega=14$ MeV&$\hbar\Omega=15.5$ MeV&$\hbar\Omega=17$ MeV\\
\hline
$E_B$($^{10}$B) & 64.75 & 63.61 & 62.78 & 61.53  \\
$Q(3^+0)$        & 8.47(6)  & 5.85  & 5.64  & 5.52   \\
$\mu(3^+0)$      & 1.80  &  1.86    & 1.85  & 1.85   \\
$\mu(1^+0)$      & 0.63(12) & 0.84  & 0.84 & 0.84   \\
$E_B$($^{10}$C) & 60.32 & 58.68 & 58.19 & 56.83  \\
$\sqrt{\langle r_p^2\rangle}$ &$2.31\pm 0.03$ &2.28 &2.21 &2.17 \\
$\delta_{\rm C} (4\hbar\Omega) [\%]$ &-&0.084 &0.091 & 0.097 \\
$\delta_{\rm C} (2\hbar\Omega) [\%]$ &-&0.055 &0.061 & 0.067
\end{tabular}
\caption{Experimental and calculated binding energies, in MeV,
magnetic moments, in $\mu_{\rm N}$, and quadrupole moments,
in $e$ fm$^2$, of $^{10}$B. Also the experimental and 
calculated binding energies, in MeV,
and the point proton radius, in fm, of $^{10}$C are presented.
The results correspond to the 4$\hbar\Omega$ calculations.
In addition, the isospin-mixing correction $\delta_{\rm C}$, in \%,
is shown as obtained both in the 4$\hbar\Omega$ calculations
and the 2$\hbar\Omega$ calculations.
Results of three different calculations with the harmonic-oscillator
parameter taken to be $\hbar\Omega=14, 15.5, 17$ MeV, respectively,
are presented.
The effective interaction used was derived from the Reid 93 
nucleon-nucleon potential. 
The experimental values are taken from Refs. 
\protect\cite{r:AS88,r:OT96}.}
\label{tab:tab1}
\end{table}

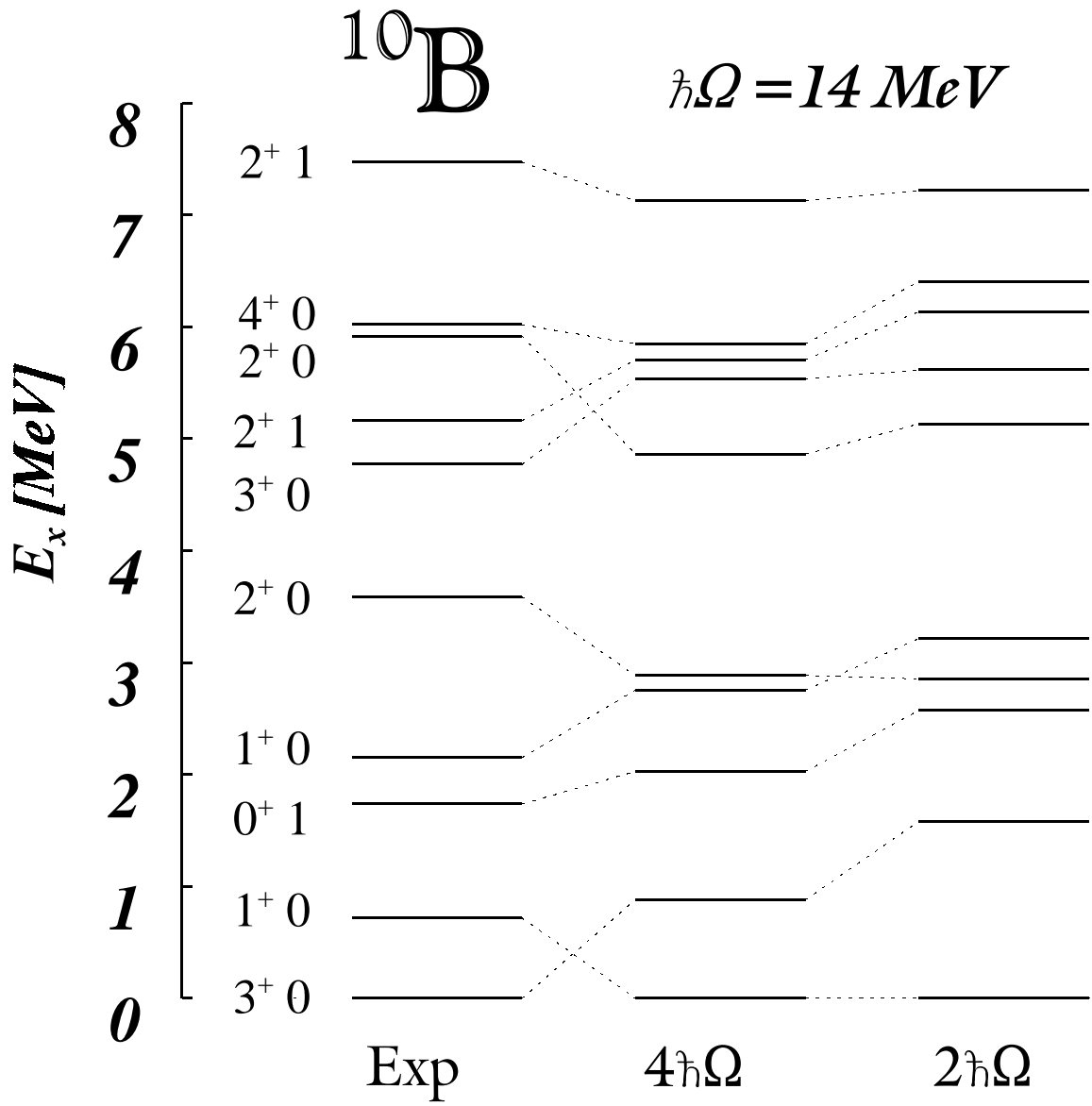
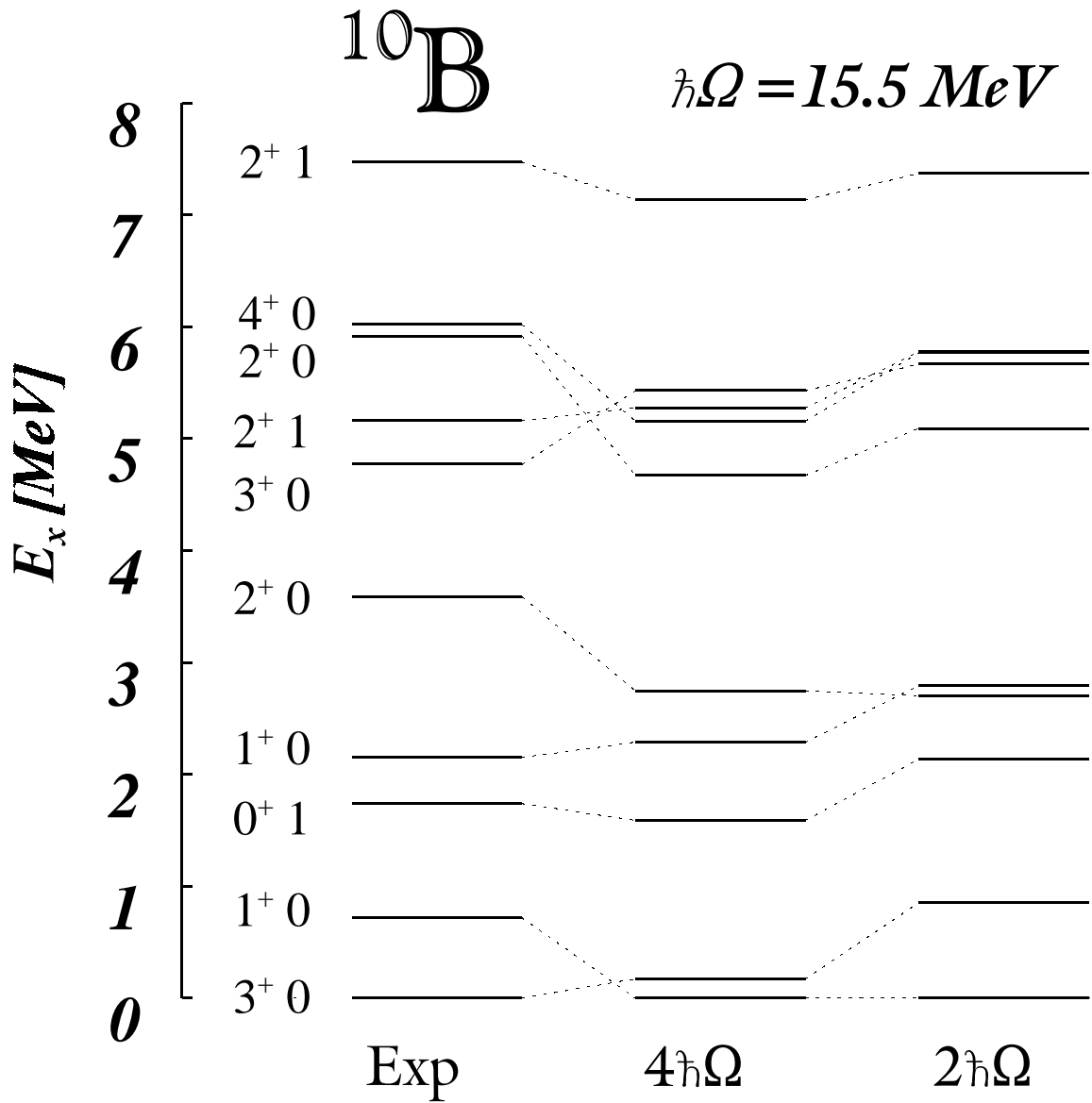
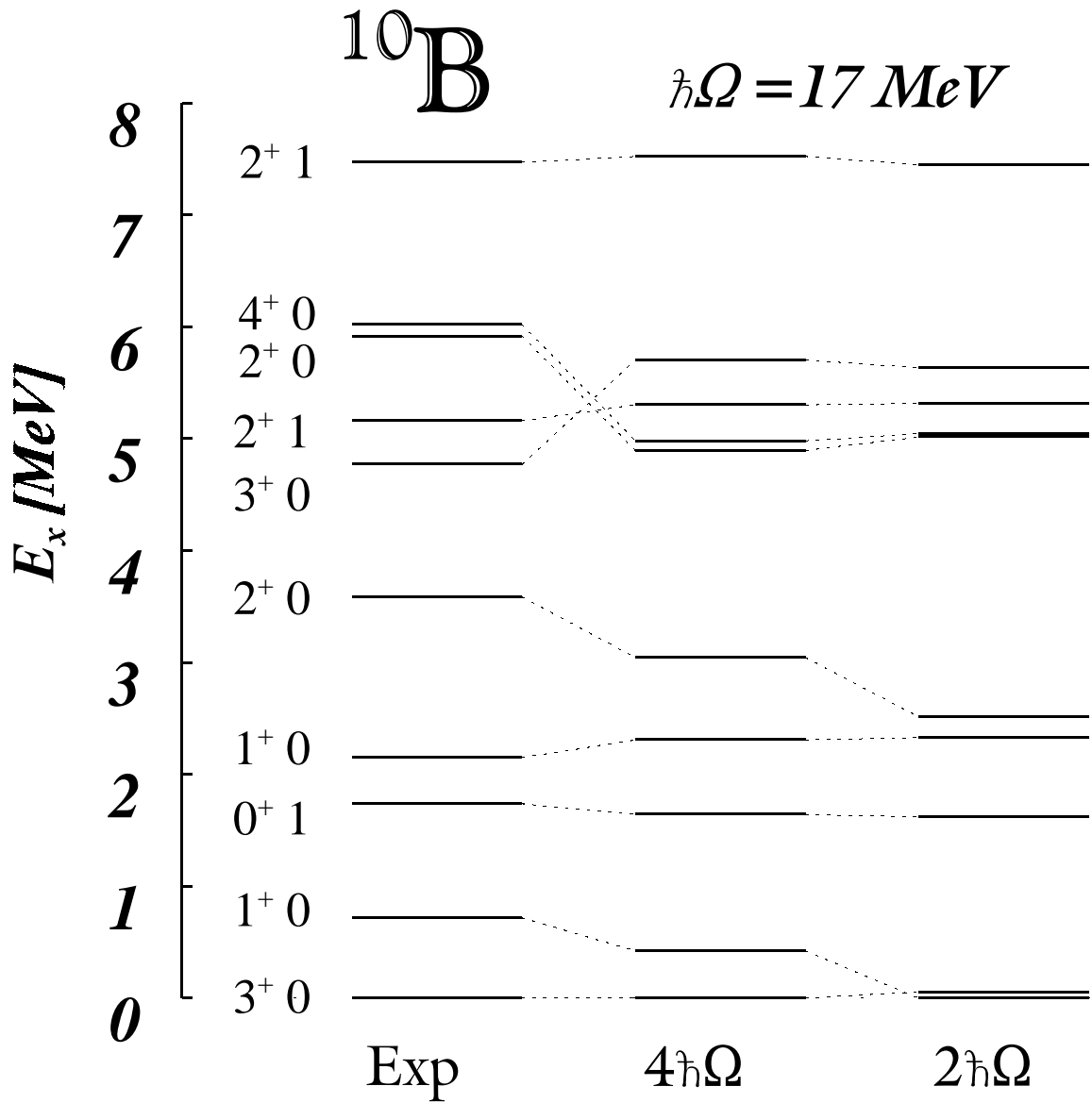
\end{document}